# On the Capacity of Fading MIMO Broadcast Channels with Imperfect Transmitter Side-Information

Amos Lapidoth*   Shlomo Shamai(Shitz)†   Michèle A. Wigger

**Abstract**

A fading broadcast channel is considered where the transmitter employs two antennas and each of the two receivers employs a single receive antenna. It is demonstrated that even if the realization of the fading is precisely known to the receivers, the high signal-to-noise (SNR) throughput is greatly reduced if, rather than knowing the fading realization *precisely*, the trasmitter only knows the fading realization *approximately*. The results are general and are not limited to memoryless Gaussian fading.

## 1 Introduction

It is quite remarkable that if the transmitter and the receivers in a multi-antenna Gaussian fading broadcast channel are cognizant of the precise realization of the fading process, then the limiting ratio, as the signal-to-noise ratio (SNR) tends to infinity, of the sum-rate capacity of the broadcast channel to the capacity of the single-user channel that results when the receivers are allowed to cooperate is one. Thus, if the fading realization is perfectly known at all terminals, then no asymptotic loss is incurred due to the lack of receiver cooperation. Here we investigate this limiting ratio in the more realistic case where the transmitter has only an approximate estimate of the fading realization. We show that the broadcast channel's sum-rate capacity suffers from this inaccuracy far more than the single-user's channel capacity. Indeed, in a broadcast channel where the transmitter employs two transmit antennas and each of the two receivers employs a single antenna, if the transmitter only has an approximate estimate of the fading realization, then this limiting ratio is upper bounded by 2/3; see Theorem 1. Thus the price of using a broadcast channel where the receivers cannot cooperate is in the lack of robustness with respect to the precision with which the transmitter knows the fading realizations.

It is interesting to note that in the single-user channel the asymptotic capacity is not sensitive to this precision at all. Indeed, the high SNR asymptotic ratio of the single-user channel capacity in the absence of any transmitter information about the fading realization to the capacity when the fading realization is known precisely is typically one.

---

*A. Lapidoth and M. Wigger are with the Signal and Information Processing Laboratory, Swiss Federal Institute of Technology (ETH) Zurich, Switzerland. E-mail: {lapidoth,wigger}@isi.ee.ethz.ch.

†S. Shamai is with the Department of Electrical Engineering, Technion—Israel Institute of Technology, Haifa, Israel. E-mail: sshlomo@ee.technion.ac.il.

It should be emphasized that throughout this paper we assume that the receivers have precise knowledge of the fading realizations. Otherwise, both the sum-rate capacity of the broadcast channel and the capacity of the single-user channel typically grow double-logarithmically [5], [9], [10] in the SNR. Also, we assume that the precision of the transmitter's estimate of the fading realization is fixed and does not improve with the SNR (which could in some scenario be motivated when only a fixed rate feedback link is available). The fading is not assumed to be Gaussian and we allow also for memory.

## 2 Previous Results

The Multi-Input-Multi-Output Gaussian Broadcast Channel (MIMO-GBC) has been studied extensively in recent years. First the optimal sum-rate has been characterized, [1], [19], [18], [22] and that has been followed by the full determination of the capacity region [20], which is one of the few examples of a non-degraded broadcast channel for which the capacity region has been determined completely. For a short review on this subject, see [2].

Most of the literature focuses on symmetric *Gaussian fading*, where the transmitter is equipped with $M$ antennas and there are $K$ single-antenna users which are statistically equivalent, that is:

$$Y = HX + N, \tag{1}$$

where $Y$ is a $K$ component vector composed of the received signal at the different $K$ users, and $H$ is a $(K, M)$ matrix with IID Gaussian components, $X$ designates the $M$ component transmit vector, and $N$ the $K$ components IID additive Gaussian noise vector.

One of the great attraction of these capacity results is the multiplexing gain, that is $\min(M, K)$ which remains the pre-log factor of the sumrate, though no cooperation of the receivers is allowed in the broadcast setting. This multiplexing gain is the central factor impacting the rate region in the high SNR regime. Though unable to provide optimal performance, numerous simple precoding techniques do preserve the multiplexing gain, and in particular techniques such as vector-perturbation precoding, as well as standard channel equalization (zero-forcing or MMSE channel inversion) procedures [14], [7], [21], [2], and references therein.

All the associated information-theoretic optimal techniques rely heavily on the precise availability of the channel propagation matrix, $H$ at both transmitter and all receivers. In fact, the underlying element in the optimal information theoretic approach to the MIMO-GBC, that is the 'Dirty-Paper' coding [1] renders the whole scheme rather sensitive to the accuracy of the CSI available at the transmitter. Also suboptimal techniques, such as the classical zero-forcing channel-equalizers assume the full availability of CSI at the transmitter [7], [21]. Already in [1], this point has been emphasized, and it has been demonstrated that for a symmetric case where all users enjoy statistically equivalent channels (as the case here), the rate region of the MIMO-GBC with no CSI available at the transmitter corresponds to that of $K = 1$. The total lack of CSI realization knowledge, causes the absolute collapse of the multiplexing gain to unity. This observation in [1] has been extended to address a general isotropic model in [5]. In [5] $H$ is assumed to be of an isotropic structure that is: $H = UG$, where $G$ is a diagonal matrix representing different channel gains to the different users, while $U$ is a unitary matrix. In this case it has been shown that the MIMO-GBC capacity region is equivalent to the capacity region associated with a single transmit antenna broadcast channel, reducing again the multiplexing gain to unity. The same conclusions remain if the $G$ matrix (i.e. gain information) is available at the transmitter site, parallelling the achievable performance in a TDMA-like regime, where the best single user is addressed [8]. It should be emphasized that this is in sharp contrast to the single user $M$ transmit $K$ received antenna case (equivalent to the MIMO-GBC, where the receivers are allowed to cooperate). In the latter, the impact of the lack of CSI is well understood and is rather limited in the high SNR regime, where optimal singular-value decomposition and water-pouring provide marginal advantage in the standard model [17], [12]. See [4] and [13] for cases where special attention is called for. In the standard single user MIMO model the full multiplexing gain is typically maintained even without any transmitter CSI.

The importance and practical implications of partial CSI knowledge, and the apparent sensitivity of the achievable rate in the MIMO-GBC to accuracy of CSI, motivated intensive research work which encompasses by now many dozens of recent contributions. Those span now different aspects, such as opportunistic approaches, linear and non-linear precoding in the presence of partial CSI, multiple beam-forming subjected to uncertainty conditions, feedback constrained signaling, Lattice reduction techniques, scheduling, antenna selection, and numerous other approaches. See [2] for a short discussion.

In [16] it was substantiated that the optimal scaling of the sumrate $C(\text{SNR})$ for MIMO-GBC, that is: $M \times \log\left(\text{SNR} \times \log(K)\right)$ can be maintained with a rather limited CSI. The limited central idea is based on transmission of random orthogonal $M$ beams and the CSI comprises the signal-to-interference ratio (SIR) observed at each beam. Transmission then is done on each beam to the user that enjoys the maximum SIR. While the $M \times \log\left(\text{SNR} \times \log(K)\right)$ conditions of fixed $M$ and SNR, while $K$ increases to infinity. The multiplexing gain however is defined as $\lim C(\text{SNR})/\log(\text{SNR})$, $\text{SNR} \to \infty$, that is fix $M$ and $K$ (though can be chosen as large as wished). Under these circumstances, it is not too difficult to realize that the opportunistic scheme in [16] is interference limited, that is: $C(\text{SNR})/\log(\text{SNR}) \to 0$. This, in principal, is also the case typifying many variants of MIMO-GBC with a variety of signaling/coding approached under finite precision CSI, that is the elements of the propagation matrix are available with a finite precision. This is not to say that any sort of partial CSI inflicts the destruction of the degrees of freedom, as is demonstrated by considering the case where perfect CSI is available just for judiciously selected subclass of scheduled users. Another example considers the availability of spatial information, say $U$ in the isotropic model discussed above. It is straightforward to realize that full multiplexing gain can be maintained if equalization, say zero forcing, is performed with respect to $U$ alone [21].

These methods however do not support finite precision CSI as is in focus here. In fact, finite precision could be associated with fixed finite rate feedback channels, as classical rate distortion theory imposes a bound on the accuracy of CSI. When the rate of the feedback channel may in fact be increased with the increase of SNR, evidently the higher the SNR, the higher the CSI precision and the qfull multiplexing gain can be preserved. In a recent contribution [6], a finite-rate feedback is considered, which scales with SNR. Multibeam forming is assumed where each user feedbacks the vector source coding of its own channel, employing unitary quantization codes. Then transmission is done to a randomly selected set of users with different beamformers. It is demonstrated that for a feedback rate proportional to $\log(\text{SNR})$ the full multiplexing gain is maintained. The resultant effect is due to the enhanced accuracy which scales exponentially with SNR (proportional to the feedback rate), hence mitigating the interference-limited regime.

## 3 Channel Model and Main Result

We consider a broadcast channel with a transmitting node $X$ and two receiving nodes $Y$ and $Z$. The transmitter employs two antennas so that the transmitted symbol takes value in $\mathbb{R}^2$. The receivers each employ a single antenna, and the received symbols thus take value in $\mathbb{R}$. Assume that the sequences of random vectors in $\mathbb{R}^2$ $\{\hat{\mathbf{A}}_k\}$, $\{\tilde{\mathbf{A}}_k\}$, $\{\hat{\mathbf{H}}_k\}$ and $\{\tilde{\mathbf{H}}_k\}$ are generated by nature according to some given law. When at time-$k$ the symbol $\mathbf{x}_k \in \mathbb{R}^2$ is sent, the received symbol at Terminal $Y$ is then given by

$$Y_k = \mathbf{A}_k^\intercal \mathbf{x}_k + N_k^{(y)} \tag{2}$$

where $\mathbf{A}_k$ models the fading and is given by

$$\mathbf{A}_k = \hat{\mathbf{A}}_k + \tilde{\mathbf{A}}_k \tag{3}$$

and where $\{N_k^{(y)}\}$ is a sequence of IID scalar zero-mean variance-$\sigma^2$ real Gaussian random variables modeling the additive noise. (Throughout this paper all Gaussian random variables are assumed to be of positive variance.) The random vector $\hat{\mathbf{A}}_k$ models the transmitter's estimate of the time-$k$ fading level and $\tilde{\mathbf{A}}_k$ denotes the transmitter's estimation error. That is we assume that the transmitter has access to the sequence $\{\hat{\mathbf{A}}_k\}$ but not to $\{\tilde{\mathbf{A}}_k\}$. The receivers, however, both have access to both $\{\hat{\mathbf{A}}_k\}$ and $\{\tilde{\mathbf{A}}_k\}$.

The received signal at Terminal $Z$ is analogously defined with similar statistical assumptions by

$$Z_k = \mathbf{H}_k^\intercal \mathbf{x}_k + N_k^{(z)} \tag{4}$$

and

$$\mathbf{H}_k = \hat{\mathbf{H}}_k + \tilde{\mathbf{H}}_k \tag{5}$$

where $\{\hat{\mathbf{H}}_k\}$ is known to both the transmitter and to the two receivers, and where $\{\tilde{\mathbf{H}}_k\}$ is only known to the two receivers. Thus, the transmitter knows $\{\hat{\mathbf{A}}_k\}$ and $\{\hat{\mathbf{H}}_k\}$ whereas both receivers know $\{\hat{\mathbf{A}}_k\}$, $\{\tilde{\mathbf{A}}_k\}$, $\{\hat{\mathbf{H}}_k\}$, and $\{\tilde{\mathbf{H}}_k\}$.

The additive noise sequences $\{N_k^{(z)}\}$ and $\{N_k^{(y)}\}$ are assumed to be independent, each being IID $\mathcal{N}(0, \sigma^2)$. Moreover, $(\{N_k^{(z)}\}, \{N_k^{(y)}\})$ is independent of $(\{\hat{\mathbf{A}}_k\}, \{\tilde{\mathbf{A}}_k\}, \{\hat{\mathbf{H}}_k\}, \{\tilde{\mathbf{H}}_k\})$ and conditional on $(\{\hat{\mathbf{A}}_k\}, \{\hat{\mathbf{H}}_k\})$ the joint law of

$$(\{N_k^{(z)}\}, \{N_k^{(y)}\}, \{\tilde{\mathbf{A}}_k\}, \{\tilde{\mathbf{H}}_k\})$$

is not depending on the input sequence $\{\mathbf{x}_k\}$.

We denote the message intended for Terminal $Y$ by $M_Y$ and assume that it is uniformly distributed over the set $\mathcal{M}_Y = \{1, \ldots, \lfloor e^{nR_Y} \rfloor\}$ where $n$ denotes the blocklength and $R_Y$ denotes the transmission rate to Terminal $Y$ in nats per channel-use. The objects $M_Z$, $\mathcal{M}_Z$, and $R_Z$ are analogously defined and we assume that $M_Y$ is independent of $M_Z$.

We consider an average power constraint

$$\frac{1}{n} \sum_{k=1}^{n} \mathsf{E}\left[\|\mathbf{X}_k(M_Y, M_Z, \hat{\mathbf{A}}_1^n, \hat{\mathbf{H}}_1^n)\|^2\right] \leq \mathcal{E}, \tag{6}$$

where $\mathbf{x}_k(m_Y, m_Z, \hat{\mathbf{a}}_1^n, \hat{\mathbf{h}}_1^n)$ denotes the symbol that is transmitted at time-$k$ when the messages $m_Y, m_Z$ are to be conveyed and when the transmitter's estimate of the fading levels are $\hat{\mathbf{a}}_1^n$ and $\hat{\mathbf{h}}_1^n$, respectively. (Here $\hat{\mathbf{A}}_1^n$ stands for $\hat{\mathbf{A}}_1, \ldots, \hat{\mathbf{A}}_n$.) We define a rate pair $(R_Y, R_Z)$ to be achievable as in the standard broadcast channel set-up [3, Ch. 14] and denote the supremum of $R_Y + R_Z$ over all achievable pairs $(R_Y, R_Z)$ by $C_{\text{Tot}}(\mathcal{E}/\sigma^2)$.

Since the capacity region of a broadcast channel depends only on the conditional marginal distributions [3, Theorem 14.61], there is no loss in generality in assuming, as we shall, that conditional on $(\{\hat{\mathbf{A}}_k\}, \{\hat{\mathbf{H}}_k\})$ the processes $\{\tilde{\mathbf{A}}_k\}$ and $\{\tilde{\mathbf{H}}_k\}$ are independent, i.e. that

$$\{\tilde{\mathbf{A}}_k\} \multimap (\{\hat{\mathbf{A}}_k\}, \{\hat{\mathbf{H}}_k\}) \multimap \{\tilde{\mathbf{H}}_k\} \tag{7}$$

forms a Markov chain.

The main result of this paper is about the asymptotic behavior of $C_{\text{Tot}}(\mathcal{E}/\sigma^2)$ as SNR tends to infinity:

**Theorem 1.** *In the above set up, assume that for every $k$ the distribution of $\mathbf{A}_k$ is the same as the distribution of $\mathbf{A}_1$ and that the distribution of $\mathbf{H}_k$ is the same as the distribution of $\mathbf{H}_1$. Moreover,*

$$\mathsf{E}\big[\|\mathbf{A}_1\|^2\big], \mathsf{E}\big[\|\mathbf{H}_1\|^2\big] < \infty. \tag{8}$$

*If, additionally,*

$$\lim_{n\to\infty} \frac{1}{n} h\big(\tilde{\mathbf{A}}_1^n \,\big|\, \hat{\mathbf{A}}_1^n, \hat{\mathbf{H}}_1^n\big) > -\infty \tag{9}$$

*and*

$$\lim_{n\to\infty} \frac{1}{n} h\big(\tilde{\mathbf{H}}_1^n \,\big|\, \hat{\mathbf{A}}_1^n, \hat{\mathbf{H}}_1^n\big) > -\infty \tag{10}$$

*then*

$$\varlimsup_{\mathcal{E}\to\infty} \frac{C_{\text{Tot}}(\mathcal{E}/\sigma^2)}{\log(1+\mathcal{E}/\sigma^2)} \leq \frac{2}{3}. \tag{11}$$

Note that since the transmitter can always choose to transmit to only one receiver, the limiting ratio is typically at least $1/2$. We conjecture that the limiting ratio is indeed $1/2$, leading to a "complete collapse of degrees of freedome" [11]. Note that in the coherent case where the transmitter knows the fading precisely this limit is typically 1. It is also typically 1, even in the absence of any transmitter side-information, if the receivers can cooperate.

## 4 Lemmas

**Lemma 1.** *For every $\Gamma > 0$ let $h_{\max}(\Gamma)$ denote the supremum of the differential entropies of all random variables $\Theta$ taking value in the interval $[-\pi, \pi)$ and satisfying*

$$\mathsf{E}\left[\log^+ \frac{1}{|\Theta|}\right] = \Gamma \tag{12}$$

*where $\log^+(\xi) \triangleq \max\{0, \log \xi\}$ for all $\xi > 0$ and $\log^+(0)$ is defined to be zero. Then*

$$-h_{\max}(\Gamma) = \Gamma - \log \Gamma - \log(2e) + o(1) \tag{13}$$

*where the $o(1)$ term tends to zero as $\Gamma \to \infty$.*

*Proof.* The proof is based on [3, Thm. 11.1.1] according to which the density $f^*(\theta)$ over $[-\pi, \pi)$ that achieves $h_{\max}(\Gamma)$ has the form

$$f^*(\theta) = \begin{cases} \frac{c}{|\theta|^\alpha} & \text{for } |\theta| \leq 1 \\ c & \text{for } 1 < |\theta| \leq \pi \end{cases} \quad (14)$$

where the constants $c > 0$ and $0 \leq \alpha < 1$ are chosen to guarantee that $f^*(\cdot)$ integrates to one over $[-\pi, \pi)$ and that the constraint is satisfied. □

**Corollary 1.** *For every $\delta > 0$ there exists some $M(\delta) > 0$ such that for any random variable $\Theta$ taking value in an interval of length $2\pi$ and having differential entropy $h(\Theta)$*

$$\mathsf{E}\left[\log^+ \frac{1}{|\Theta|}\right] \leq \max\{M(\delta), -(1+\delta)h(\Theta)\}. \quad (15)$$

*Proof.* We first note that it suffices to prove the bound for the case where $\Theta$ takes value in the symmetric interval $[-\pi, \pi)$. Indeed, if $\Theta$ takes value in any (not necessarily symmetric about the origin) length-$2\pi$ interval then

$$\mathsf{E}\left[\log^+ \frac{1}{|\Theta|}\right] \leq \mathsf{E}\left[\log^+ \frac{1}{|\tilde{\Theta}|}\right]$$

where

$$\tilde{\Theta} = \Theta \mod [-\pi, \pi).$$

But $h(\tilde{\Theta}) = h(\Theta)$ so that the result will follow from the result for $\tilde{\Theta}$, which takes value in $[-\pi, \pi)$.

We now proceed to prove the result for the case where $\Theta$ takes value in $[-\pi, \pi)$. By Lemma 1 there exists some $M(\delta)$ such that

$$-h_{\max}(\Gamma) \geq \frac{1}{1+\delta}\Gamma, \quad \Gamma \geq M(\delta) \quad (16)$$

and hence,

$$\Gamma \leq \max\{M(\delta), -(1+\delta)h_{\max}(\Gamma)\}. \quad (17)$$

The corollary now follows from this inequality by noting that, by the definition of $h_{\max}$ in the lemma, if $\Theta \in [-\pi, \pi)$ is arbitrary (i.e., not necessarily having the max-entropy distribution) then

$$h(\Theta) \leq h_{\max}\left(\mathsf{E}\left[\log^+ \frac{1}{|\Theta|}\right]\right). \quad (18)$$

□

**Lemma 2.** *Let the random variable $X$ be of finite second moment and independent of the Gaussian random variable of positive variance $U$. Let $F_S(\cdot)$ and $F_T(\cdot)$ be two distribution functions on the real line with finite variances. Then for any distribution function $F(\cdot, \cdot)$ on $\mathbb{R}^2$ of marginals $F(\cdot, +\infty) = F_S(\cdot)$ and $F(+\infty, \cdot) = F_T(\cdot)$ we have*

$$\int_{-\infty}^{\infty} h(sX + U) \dot{F}_S(s) \geq \int_{-\infty}^{\infty} h(tX + U) \dot{F}_T(t) - \int_{-\infty}^{\infty}\int_{-\infty}^{\infty} \log^+ \frac{|t|}{|s|} \dot{F}(s, t). \quad (19)$$

*Consequently, using $\log^+(a/b) \leq \log^+(a) + \log^+(1/b)$, we have*

$$\int_{-\infty}^{\infty} h(sX+U)\dot{F}_S(s) \geq \int_{-\infty}^{\infty} h(tX+U)\dot{F}_T(t) - \int_{-\infty}^{\infty} \log^+ |t|\dot{F}_T - \int_{-\infty}^{\infty} \log^+ \frac{1}{|s|}\dot{F}_S(s). \quad (20)$$

*Proof.* The proof is based on two inequalities:

$$h(sX + U) \geq h(tX + U) - \log \frac{|t|}{|s|}, \quad |t| \geq |s| \tag{21}$$

and

$$h(sX + U) \geq h(tX + U), \quad |t| \leq |s| \tag{22}$$

which combine to prove that

$$h(sX + U) \geq h(tX + U) - \log^+ \frac{|t|}{|s|} \tag{23}$$

from which the lemma follows by integration.

Both inequalities follow by noting that a zero-mean variance-$(a^2 + b^2)$ random variable can be expressed as the sum of two independent zero-mean Gaussians of respective variances $a^2$ and $b^2$. Indeed, (21) follows from:

$$\begin{aligned} h(sX + U) &= h\left(\frac{t}{s}(sX + U)\right) - \log \frac{|t|}{|s|} \\ &= h\left(tX + \frac{t}{s}U\right) - \log \frac{|t|}{|s|} \\ &\geq h(tX + U) - \log \frac{|t|}{|s|}, \quad \frac{|t|}{|s|} \geq 1 \end{aligned}$$

where the first equality follows from the behavior of differential entropy under scaling and the last inequality follows by writing $(t/s) \cdot U$ as a sum of two independent Gaussians of variances $\sigma^2$ and $(t^2/s^2 - 1)\sigma^2$ and by conditioning on the latter.

Similarly, (22) follows for $|t| \leq |s|$ using the Data Processing Inequality:

$$\begin{aligned} I(X; sX + U) &\geq I\left(X; \frac{t}{s}(sX + U)\right) \\ &= I\left(X; tX + \frac{t}{s}U\right) \\ &\geq I(X; tX + U), \quad \frac{|t|}{|s|} \leq 1 \end{aligned}$$

from which (22) follows by expanding mutual information in terms of differential entropies. Here the first inequality follows because scaling does not change mutual information unless the scaling is by zero, and the last inequality follows by noting that $U$ can be written as a sum of independent Gaussians of variances $(t^2/s^2)\sigma^2$ and $(1 - t^2/s^2)\sigma^2$ and by invoking the Data Processing Inequality. □

**Lemma 3.** *Let the random vector $\mathbf{W}$ take value in $\mathbb{R}^2$ and let $R$ and $\Theta$ denote its magnitude and phase in the sense that $R = \|\mathbf{W}\|$, $\Theta \in [-\pi, \pi)$ and $\mathbf{W}^\mathsf{T} = (R\cos\Theta, R\sin\Theta)$. Assume that $\mathbf{W}$ is of finite second moment and finite differential entropy. Then*

$$\begin{aligned} h(\Theta) &= h(W) - h(R|\Theta) - \mathsf{E}[\log R] \tag{24} \\ &\geq h(W) - h(R) - \mathsf{E}[\log R] \tag{25} \end{aligned}$$

*with equality if, and only if, $R$ and $\Theta$ are independent.*

*Proof.* The equality follows directly from the behavior of differential entropy under change of coordinates; see, for example, [10, Lemma 6.16]. The inequality follows because conditioning can only decrease differential entropy unless the random variables are independent. □

**Lemma 4.** *Let the pair $(X, Y)$ of finite-variance random variables be independent of the pair $(U, V)$ where $U$ and $V$ are IID $\mathcal{N}(0, \sigma^2)$ for some $\sigma^2 > 0$. Let*

$$\mathcal{H}(\theta) \triangleq h\big((X+U)\cos\theta + (Y+V)\sin\theta\big), \quad -\pi \leq \theta < \pi,$$

$$\mathcal{H}_{\sup} \triangleq \sup_{-\pi \leq \theta < \pi} \mathcal{H}(\theta).$$

*Then for any distribution function $F_\Theta(\cdot)$ on $[-\pi, \pi)$,*

$$\int_{-\pi}^{\pi} \mathcal{H}(\theta) F_\Theta(\theta) \geq \frac{1}{2}\mathcal{H}_{\sup} + \frac{1}{4}\log(2\pi e\sigma^2) + \inf_{-\pi \leq \phi < \pi} \int_{-\pi}^{\pi} \log|\sin(\theta-\phi)| F_\Theta(\theta) \quad (26)$$

$$\geq \frac{1}{2}\mathcal{H}_{\sup} + \frac{1}{4}\log(2\pi e\sigma^2) - \log\frac{\pi}{2} - 3\max\left\{M(1/2), -\frac{3}{2}h(\Theta)\right\} \quad (27)$$

*where $M(1/2)$ is a universal constant that is defined in Corollary 1. Using $h(\Theta) \leq \log(2\pi)$ we have*

$$\max\{M(1/2), -\frac{3}{2}h(\Theta) \leq M(1/2) - \frac{3}{2}h(\Theta) + \frac{3}{2}\log 2\pi \quad (28)$$

*and hence*

$$\int_{-\pi}^{\pi} \mathcal{H}(\theta) F_\Theta(\theta) \geq \frac{1}{2}\mathcal{H}_{\sup} + \frac{1}{4}\log\sigma^2 + \frac{9}{2}h(\Theta) - \gamma, \quad (29)$$

*where $\gamma$ is some universal constant.*

*Proof.* Define $\mathcal{H}_{\inf} = \inf_{\theta \in [-\pi, \pi)} \mathcal{H}(\theta)$ and note that $\mathcal{H}_{\inf} \geq 1/2\log(2\pi e\sigma^2)$. Also, because $X$ and $Y$ have finite second moments, $\mathcal{H}_{\sup} < \infty$. For some fixed $\delta > 0$ let $\theta_{\max}$ and $\theta_{\min}$ be such that

$$\mathcal{H}(\theta_{\min}) < \mathcal{H}_{\inf} + \delta, \qquad \mathcal{H}(\theta_{\max}) > \mathcal{H}_{\sup} - \delta. \quad (30)$$

For any $\theta_1, \theta_2 \in [-\pi, \pi)$ denote by $J(\theta_1, \theta_2)$ the joint differential entropy

$$J(\theta_1, \theta_2) \triangleq h\big(\cos\theta_1(X+U) + \sin\theta_1(Y+V)\big), \cos\theta_2(X+U) + \sin\theta_2(Y+V)\big). \quad (31)$$

Note that

$$J(\theta_1, \theta_2) \leq \mathcal{H}(\theta_1) + \mathcal{H}(\theta_2) \quad (32)$$

and that because

$$\begin{pmatrix} \cos\theta_1(X+U) + \sin\theta_1(Y+V) \\ \cos\theta_2(X+U) + \sin\theta_2(Y+V) \end{pmatrix} = \begin{pmatrix} \cos\theta_1 & \sin\theta_1 \\ \cos\theta_2 & \sin\theta_2 \end{pmatrix} \begin{pmatrix} X+U \\ Y+V \end{pmatrix}$$

it follows that

$$J(\theta_1, \theta_2) = J(0, \pi/2) + \log|\sin(\theta_2 - \theta_1)|. \quad (33)$$

We now have for any $\theta \in [-\pi, \pi)$

$$2\mathcal{H}(\theta) \geq \mathcal{H}(\theta) + \mathcal{H}(\theta_{\min}) - \delta$$
$$\geq J(\theta, \theta_{\min}) - \delta$$
$$= J(0, \pi/2) + \log|\sin(\theta - \theta_{\min})| - \delta$$
$$= J(\theta_{\max}, \theta_{\max} + \pi/2) + \log|\sin(\theta - \theta_{\min})| - \delta$$
$$\geq \mathcal{H}(\theta_{\max}) + \frac{1}{2}\log(2\pi e\sigma^2) + \log|\sin(\theta - \theta_{\min})| - \delta$$
$$\geq \mathcal{H}_{\sup} + \frac{1}{2}\log(2\pi e\sigma^2) + \log|\sin(\theta - \theta_{\min})| - 2\delta$$

from which (26) follows by letting $\delta$ tend to zero and integrating over $\theta$. Here the first inequality follows from the definition of $\mathcal{H}(\theta_{\min})$; the subsequent inequality from (32); the two subsequent equalities from (33); the subsequent inequality by the chain rule and since conditioning reduces entropy and then using that $U$ and $V$ are independent Gaussians; and the final inequality by the definition of $\theta_{\max}$.

To prove (27) we use the inequalities

$$|\sin\xi| \geq \min\left\{\frac{2}{\pi}|\xi|, \frac{2}{\pi}|\xi - \pi|, \frac{2}{\pi}|\xi + \pi|\right\}, \quad -\pi \leq \xi < \pi \qquad (34)$$

and

$$\log\min\{a, b, c\} \geq \log^- a + \log^- b + \log^- c, \quad a, b, c \geq 0$$

where $\log^-(a) \triangleq -\log^+(1/a)$ is given by $\min\{\log a, 0\}$ for $a > 0$ and as $-\infty$ for $a = 0$ to obtain

$$-\log|\sin(\theta - \phi)| \leq \log\frac{\pi}{2} + \log^+\frac{1}{|\theta - \phi|} + \log^+\frac{1}{|\theta - \phi + \pi|} + \log^+\frac{1}{|\theta - \phi - \pi|}.$$

From here (27) can be derived from (26) as follows:

$$\inf_{-\pi \leq \phi < \pi}\int_{-\pi}^{\pi}\log|\sin(\theta - \phi)|\mathrm{F}_\Theta(\theta)$$

$$= -\sup_{-\pi \leq \phi < \pi}\int_{-\pi}^{\pi}-\log|\sin(\theta - \phi)|\mathrm{F}_\Theta(\theta)$$

$$\geq -\sup_{-\pi \leq \phi < \pi}\int_{-\pi}^{\pi}\left\{\log\frac{\pi}{2} + \log^+\frac{1}{|\theta - \phi|} + \log^+\frac{1}{|\theta - \phi + \pi|} + \log^+\frac{1}{|\theta - \phi - \pi|}\right\}\mathrm{F}_\Theta(\theta)$$

$$\geq \log\frac{\pi}{2} - 3\sup_{\alpha \in \mathbb{R}}\mathsf{E}\left[\log^+\frac{1}{|\Theta - \alpha|}\right] \qquad (35)$$

from which (27) follows using Corollary 1 because for any $\alpha \in \mathbb{R}$ the random variable $\Theta - \alpha$ takes value in an interval of length $2\pi$ and has differential entropy $h(\Theta)$. □

**Lemma 5.** *Let the random vectors $\mathbf{A}$ and $\mathbf{H}$ each take value in $\mathbb{R}^2$ independently of the zero-mean variance-$\sigma^2$ ($\sigma^2 > 0$) random variable $U$. Let $\|\mathbf{A}\| \geq 0$ and $\Theta_A \in [-\pi, \pi)$ denote the magnitude and angle of $\mathbf{A}$ in the sense that $\mathbf{A}^\mathsf{T} = (\|\mathbf{A}\|\cos\Theta_A, \|\mathbf{A}\|\sin\Theta_A)$. Let $\mathbf{X} \in \mathbb{R}^2$ be independent of $(\mathbf{A}, \mathbf{H}, U)$ and assume that $\mathsf{E}[\|\mathbf{A}\|^2]$, $\mathsf{E}[\|\mathbf{H}\|^2]$ and $\mathsf{E}[\|\mathbf{X}\|^2]$ are all finite. Then,*

$$h(\mathbf{A}^\mathsf{T}\mathbf{X} + U|\mathbf{A}) \geq \frac{1}{2}h(\mathbf{H}^\mathsf{T}\mathbf{X} + U|\mathbf{H})$$
$$-\frac{1}{2}\mathsf{E}\left[\log^+\|\mathbf{H}\|\right] - \mathsf{E}\left[\log^+\frac{1}{\|\mathbf{A}\|}\right]$$
$$+\frac{1}{4}\log\sigma^2 - \gamma + \frac{9}{2}h(\Theta_\mathbf{A}). \qquad (36)$$

*Moreover, if additionally $h(\mathbf{A}) > -\infty$ then*

$$h(\mathbf{A}^\mathsf{T}\mathbf{X} + U|\mathbf{A}) \geq \frac{1}{2}h(\mathbf{H}^\mathsf{T}\mathbf{X} + U|\mathbf{H}) - \frac{1}{2}\mathsf{E}\left[\log^+\|\mathbf{H}\|\right] - \mathsf{E}\left[\log^+\frac{1}{\|\mathbf{A}\|}\right]$$
$$+\frac{1}{4}\log\sigma^2 - \gamma + \frac{9}{2}\left(h(\mathbf{A}) - h(\|\mathbf{A}\|) - \frac{1}{2}\mathsf{E}\left[\log\|\mathbf{A}\|^2\right]\right) \qquad (37)$$

*where $\gamma$ is a universal constant.*

*Proof.* We first use polar coordinates to write

$$h(\mathbf{A}^\mathsf{T}\mathbf{X} + U|\mathbf{A}) \triangleq \int h(\mathbf{a}^\mathsf{T}\mathbf{X} + U) F_\mathbf{A}(\mathbf{a})$$

$$= \int\int h\big(r(\cos\theta X[1] + \sin\theta X[2]) + U\big) F_{\|\mathbf{A}\| \,|\, \Theta_\mathbf{A}}(r|\theta) F_{\Theta_A}(\theta).$$

Conditional on $\Theta_A = \theta$ we now apply Lemma 2 with the substitution of $\cos\theta X[1] + \sin\theta X[2]$ for $X$; with the substitution of $\|\mathbf{A}\|$ (of law $F_{\|\mathbf{A}\| \,|\, \Theta_A = \theta}$) for $S$; and with the substition of 1 for $T$ to obtain

$$h(\mathbf{A}^\mathsf{T}\mathbf{X} + U|\mathbf{A})$$
$$\geq \int \left\{ h(\cos\theta X[1] + \sin\theta X[2] + U) - \int_\infty^\infty \log^+ \frac{1}{\|\mathbf{a}\|} F_{\|\mathbf{A}\|\,|\,\Theta_A=\theta}(\|\mathbf{a}\|) \right\} F_{\Theta_\mathbf{A}}(\theta)$$
$$= \int h(\cos\theta X[1] + \sin\theta X[2] + U) F_{\Theta_\mathbf{A}}(\theta) - \mathsf{E}\left[\log^+ \frac{1}{\|\mathbf{A}\|}\right]. \tag{38}$$

We similarly express $h(\mathbf{H}^\mathsf{T}\mathbf{X} + U|\mathbf{H})$ using polar coordinates and conditional on $\Theta_H = \theta$ we apply Lemma 2 with the substitution of $\cos\theta X[1] + \sin\theta X[2]$ for $X$; with the substitution of 1 for $S$; and with the substitution of $\|\mathbf{H}\|$ (of law $F_{\|\mathbf{H}\|\,|\,\Theta_H=\theta}$) for $T$ to obtain

$$h(\cos\theta X[1] + \sin\theta X[2] + U) \geq h(\mathbf{H}^\mathsf{T}\mathbf{X} + U|\mathbf{H}) - \mathsf{E}\left[\log^+ \|\mathbf{H}\|\right]. \tag{39}$$

Inequality (36) now follows from (38) and by applying (29) of Lemma 4 and by noting that

$$\sup_{|\theta|\leq\pi} h(\cos\theta X[1] + \sin\theta X[2] + U) \geq \int h(\cos\theta X[1] + \sin\theta X[2] + U) F_{\Theta_\mathbf{H}}(\theta)$$

and finally applying (39). Inequality (37) follows from (36) using Lemma 3. □

**Corollary 2.** *Let $\mathbf{X}$, $\mathbf{A}$ and $\mathbf{H}$ take value in $\mathbb{R}^2$ with $\mathbf{X}$—◦—$S$—◦—$\mathbf{A}$ and $\mathbf{X}$—◦—$S$—◦—$\mathbf{H}$ forming Markov chains and $U$ being $\mathcal{N}(0,\sigma^2)$ distributed and independent of $(\mathbf{X}, S, \mathbf{A}, \mathbf{H})$. If $\mathsf{E}[\|\mathbf{X}\|^2], \mathsf{E}[\|\mathbf{A}\|^2], \mathsf{E}[\|\mathbf{H}\|^2] < \infty$ and $h(\mathbf{A}|S) > -\infty$, then*

$$h(\mathbf{A}^\mathsf{T}\mathbf{X} + U|\mathbf{A}, S) \geq \frac{1}{2}h(\mathbf{H}^\mathsf{T}\mathbf{X} + U|\mathbf{H}, S) - \frac{1}{2}\mathsf{E}\left[\log^+ \|\mathbf{H}\|\right] - \mathsf{E}\left[\log^+ \frac{1}{\|\mathbf{A}\|}\right]$$
$$-\gamma + \frac{1}{4}\log\sigma^2 + \frac{9}{2}\big(h(\mathbf{A}|S) - h(\|\mathbf{A}\|\,|\,S) - \frac{1}{2}\mathsf{E}\left[\log\|\mathbf{A}\|^2\right]\big) \tag{40}$$
$$\geq \frac{1}{2}h(\mathbf{H}^\mathsf{T}\mathbf{X} + U|\mathbf{H}, S) - \frac{1}{2}\mathsf{E}\left[\log^+\|\mathbf{H}\|\right] - \mathsf{E}\left[\log^+ \frac{1}{\|\mathbf{A}\|}\right] - \gamma$$
$$+\frac{1}{4}\log\sigma^2 + \frac{9}{2}\big(h(\mathbf{A}|S) - \frac{1}{2}\log(2\pi e \mathsf{E}[\|\mathbf{A}\|^2]) - \frac{1}{2}\log\mathsf{E}[\|\mathbf{A}\|^2]\big) \tag{41}$$

*where $\gamma$ is, as in the lemma, a universal constant.*

*Proof.* Inequality (40) follows directly from the lemma by conditioning on $S$. Inequality (41) follows from (40) by upper bounding $h(\|\mathbf{A}\|\,|\,S)$ with $h(\|\mathbf{A}\|)$ and by upper bounding the latter by the differential entropy of a Gaussian of equal second moment. □

**Lemma 6.** *Let $X$ and $S$ be possibly dependent random variables with $X$ satisfying $\mathsf{E}[X^2] \leq \mathcal{E}$ and $S$ having a finite second moment. Then*

$$\frac{1}{2}\mathsf{E}\left[\log\big(2\pi e(S^2 \mathsf{E}[X^2 \mid S] + \sigma^2)\big)\right] \leq \frac{1}{2}\log\big(2\pi e(\mathsf{E}[S^2]\mathcal{E} + \sigma^2)\big) + \Delta \tag{42}$$

*where $\Delta$ is upper bounded by $\log(e)/e$.*

*Proof.* This is a fairly standard result in "optimal power allocation" [3, Sec. 10.4 & 10.5]. The maximum of the left hand side (LHS) of (42) over all choices for $\mathsf{E}[X^2 \mid S]$ integrating to $\mathcal{E}$ has been extensively studied in the literature. Schein & Trott [15] and Zamir & Erez [23] have, for example, shown that the choice $\mathsf{E}[X^2 \mid S] = \mathcal{E}$ can be off from the optimum by at most $\log(e)/e$ from which the firm bound follows using Jensen's inequality. □

## 5 Proof of the Theorem

Assume a sequence of encoding and decoding functions for the setting described in Section 3 of rates $R_Y$ and $R_Z$ such that the average probabilities of errors as defined in the standard broadcast set-up tend to 0. Given such a block-length $n$ encoder define $Y_1^n$ and $Z_1^n$ to be the $n$-length vectors of the outputs $Y_1, \ldots, Y_n$ and $Z_1, \ldots, Z_n$, respectively. Similarly define $\tilde{\mathbf{A}}_1^n, \tilde{\mathbf{H}}_1^n, \hat{\mathbf{A}}_1^n$ and $\hat{\mathbf{H}}_1^n$ to be the $n$-length sequences of fading coefficients. Using Fano's Inequality [3] and assuming that Terminal $Z$ knows the message $M_Y$ intended for Terminal $Y$, one obtains an upper bound on the rates

$$\begin{aligned} R_Y &< \frac{1}{n} I(M_Y; Y_1^n | \tilde{\mathbf{A}}_1^n, \hat{\mathbf{A}}_1^n, \hat{\mathbf{H}}_1^n) + \epsilon(n), \\ R_Z &< \frac{1}{n} I(M_Z; Z_1^n | M_Y, \tilde{\mathbf{H}}_1^n, \hat{\mathbf{A}}_1^n, \hat{\mathbf{H}}_1^n) + \epsilon(n), \end{aligned} \quad (43)$$

where the term $\epsilon(n)$ tends to 0 as $n$ tends to infinity.

The rate $R_Y$ can be further upper bounded using the following lower bound

$$h(Y_1^n|M_Y, \tilde{\mathbf{A}}_1^n, \hat{\mathbf{A}}_1^n, \hat{\mathbf{H}}_1^n)$$
$$= \sum_{i=1}^n \left\{ h(Y_1^i, Z_{i+1}^n|M_Y, \tilde{\mathbf{A}}_1^i, \tilde{\mathbf{H}}_{i+1}^n, \hat{\mathbf{A}}_1^n, \hat{\mathbf{H}}_1^n) - h(Y_1^{i-1}, Z_i^n|M_Y, \tilde{\mathbf{A}}_1^{i-1}, \tilde{\mathbf{H}}_i^n, \hat{\mathbf{A}}_1^n, \hat{\mathbf{H}}_1^n) \right\}$$
$$+ h(Z_1^n|M_Y, \tilde{\mathbf{H}}_1^n, \hat{\mathbf{A}}_1^n, \hat{\mathbf{H}}_1^n)$$
$$= \sum_{i=1}^n \Big\{ h(Y_i|Y_1^{i-1}, Z_{i+1}^n, M_Y, \tilde{\mathbf{A}}_1^i, \tilde{\mathbf{H}}_{i+1}^n, \hat{\mathbf{A}}_1^n, \hat{\mathbf{H}}_1^n)$$
$$- h(Z_i|Y_1^{i-1}, Z_{i+1}^n, M_Y, \tilde{\mathbf{A}}_1^{i-1}, \tilde{\mathbf{H}}_i^n, \hat{\mathbf{A}}_1^n, \hat{\mathbf{H}}_1^n) \Big\} + h(Z_1^n|M_Y, \tilde{\mathbf{H}}_1^n, \hat{\mathbf{A}}_1^n, \hat{\mathbf{H}}_1^n) \quad (44)$$
$$\geq \sum_{i=1}^n \left\{ -\frac{1}{2} h(Z_i|Y_1^{i-1}, Z_{i+1}^n, M_Y, \tilde{\mathbf{A}}_1^{i-1}, \tilde{\mathbf{H}}_i^n, \hat{\mathbf{A}}_1^n, \hat{\mathbf{H}}_1^n) + \frac{1}{4} \log \sigma^2 - \gamma - \frac{1}{2} \mathsf{E}\left[ \log^+ \|\hat{\mathbf{H}}_i + \tilde{\mathbf{H}}_i\| \right] \right.$$
$$- \mathsf{E}\left[ \log^+ \frac{1}{\|\hat{\mathbf{A}}_i + \tilde{\mathbf{A}}_i\|} \right] + \frac{9}{2}\Big( h(\hat{\mathbf{A}}_i + \tilde{\mathbf{A}}_i|Y_1^{i-1}, Z_{i+1}^n, M_Y, \tilde{\mathbf{A}}_1^{i-1}, \tilde{\mathbf{H}}_{i+1}^n, \hat{\mathbf{A}}_1^n, \hat{\mathbf{H}}_1^n)$$
$$\left. -\frac{1}{2}\log\Big(2\pi e \mathsf{E}\big[\|\hat{\mathbf{A}}_i + \tilde{\mathbf{A}}_i\|^2\big]\Big) - \frac{1}{2}\log \mathsf{E}\big[\|\hat{\mathbf{A}}_i + \tilde{\mathbf{A}}_i\|^2\big] \Big) \right\}$$
$$+ h(Z_1^n|M_Y, \tilde{\mathbf{H}}_1^n, \hat{\mathbf{A}}_1^n, \hat{\mathbf{H}}_1^n) \quad (45)$$
$$\geq \sum_{i=1}^n \left\{ -\frac{1}{2} h(Z_i|Z_{i+1}^n, M_Y, \tilde{\mathbf{H}}_1^n, \hat{\mathbf{A}}_1^n, \hat{\mathbf{H}}_1^n) - \frac{1}{2}\mathsf{E}\left[\log^+ \|\hat{\mathbf{H}}_i + \tilde{\mathbf{H}}_i\|\right] - \mathsf{E}\left[\log^+ \frac{1}{\|\hat{\mathbf{A}}_i + \tilde{\mathbf{A}}_i\|}\right] \right.$$
$$+ \frac{1}{4}\log\sigma^2 - \gamma + \frac{9}{2}\Big( h(\tilde{\mathbf{A}}_i|\tilde{\mathbf{A}}_1^{i-1}, \hat{\mathbf{A}}_1^n, \hat{\mathbf{H}}_1^n) - \frac{1}{2}\log\Big(2\pi e \mathsf{E}\big[\|\hat{\mathbf{A}}_i + \tilde{\mathbf{A}}_i\|^2\big]\Big)$$
$$\left. - \frac{1}{2}\log \mathsf{E}\big[\|\hat{\mathbf{A}}_i + \tilde{\mathbf{A}}_i\|^2\big]\Big) \right\} + h(Z_1^n|M_Y, \tilde{\mathbf{H}}_1^n, \hat{\mathbf{A}}_1^n, \hat{\mathbf{H}}_1^n) \quad (46)$$
$$\geq \frac{1}{2} h(Z_1^n|M_Y, \tilde{\mathbf{H}}_1^n, \hat{\mathbf{A}}_1^n, \hat{\mathbf{H}}_1^n) + \frac{n}{4}\log\sigma^2 - n\gamma - \frac{9n}{4}\log(2\pi e) + \frac{9}{2} h(\tilde{\mathbf{A}}_1^n|\hat{\mathbf{A}}_1^n, \hat{\mathbf{H}}_1^n)$$
$$- \frac{n}{2}\mathsf{E}\left[\log^+ \|\hat{\mathbf{H}}_1 + \tilde{\mathbf{H}}_1\|\right] - n\mathsf{E}\left[\log^+ \frac{1}{\|\hat{\mathbf{A}}_1 + \tilde{\mathbf{A}}_1\|}\right] - \frac{9n}{2}\log \mathsf{E}\big[\|\hat{\mathbf{A}}_1 + \tilde{\mathbf{A}}_1\|^2\big] \quad (47)$$

where $\gamma$ is a universal constant. Here Equation (44) is obtained by subtracting the term $\sum_{i=1}^n h(Y_1^{i-1} Z_{i+1}^n|M_Y, \tilde{\mathbf{A}}_1^i, \tilde{\mathbf{H}}_{i+1}^n, \hat{\mathbf{A}}_1^n, \hat{\mathbf{H}}_1^n)$ and then adding $\sum_{i=1}^n h(Y_1^{i-1} Z_{i+1}^n|M_Y, \tilde{\mathbf{A}}_1^{i-1}, \tilde{\mathbf{H}}_i^n, \hat{\mathbf{A}}_1^n, \hat{\mathbf{H}}_1^n)$ where both sums are equal because $(\tilde{\mathbf{A}}_i, \tilde{\mathbf{H}}_i)$—$\circ$—$(M_Y, \tilde{\mathbf{A}}_1^{i-1}, \tilde{\mathbf{H}}_{i+1}^n, \hat{\mathbf{A}}_1^n, \hat{\mathbf{H}}_1^n)$—$\circ$—$(Y_1^{i-1}, Z_{i+1}^n)$ forms a Markov chain. In the following let $S_i = (Y_1^{i-1}, Z_{i+1}^n, M_Y, \tilde{\mathbf{A}}_1^{i-1}, \tilde{\mathbf{H}}_{i+1}^n, \hat{\mathbf{A}}_1^n, \hat{\mathbf{H}}_1^n)$. Then the Markov chains $(\hat{\mathbf{A}}_i + \tilde{\mathbf{A}}_i)$—$\circ$—$S_i$—$\circ$—$\mathbf{X}_i$ and $(\hat{\mathbf{H}}_i + \tilde{\mathbf{H}}_i)$—$\circ$—$S_i$—$\circ$—$\mathbf{X}_i$ are fulfilled and $h(\hat{\mathbf{A}}_i + \tilde{\mathbf{A}}_i|S_i) > -\infty$ holds due to (9) and the Markov Chain $\tilde{\mathbf{A}}_i$—$\circ$—$\left(\tilde{\mathbf{A}}_1^{i-1}, \hat{\mathbf{A}}_1^n, \hat{\mathbf{H}}_1^n\right)$—$\circ$—$S_i$. Thus Corollary 2 can be applied and Inequality (45) follows. In Inequality (46) we used that differential entropy cannot be increased by conditioning and is invariant under shift together with the fact that $Z_i$—$\circ$—$(S_i, \tilde{\mathbf{H}}_i)$—$\circ$—$\tilde{\mathbf{H}}_1^{i-1}$ and $\tilde{\mathbf{A}}_i$—$\circ$—$(\tilde{\mathbf{A}}_1^{i-1}, \hat{\mathbf{A}}_1^n, \hat{\mathbf{H}}_1^n)$—$\circ$—$S_i$ form Markov chains. Finally, Inequality (47) follows due to the chain rule and because the marginal distributions of $\mathbf{A}_i$ and $\mathbf{H}_i$ do not depend on the index $i$. Note that the Markov chains used to justify these inequalities follow mainly from Assumption (7) and because the transmitter has access only to $\hat{\mathbf{A}}_1^n$ and $\hat{\mathbf{H}}_1^n$ but not to $\tilde{\mathbf{A}}_1^n$ or to $\tilde{\mathbf{H}}_1^n$.

From this follows that

$$R_Y < \frac{1}{n}h(Y_1^n|\tilde{\mathbf{A}}_1^n, \hat{\mathbf{A}}_1^n, \hat{\mathbf{H}}_1^n) - \frac{1}{2n}h(Z_1^n|M_Y, \tilde{\mathbf{H}}_1^n, \hat{\mathbf{A}}_1^n, \hat{\mathbf{H}}_1^n) - \frac{9}{2n}h(\tilde{\mathbf{A}}_1^n|\hat{\mathbf{A}}_1^n, \hat{\mathbf{H}}_1^n) - \frac{1}{4}\log\sigma^2$$

$$+ \frac{1}{2}\mathsf{E}\left[\log^+ \|\hat{\mathbf{H}}_1 + \tilde{\mathbf{H}}_1\|\right] + \mathsf{E}\left[\log^+ \frac{1}{\|\hat{\mathbf{A}}_1 + \tilde{\mathbf{A}}_1\|}\right] + \gamma$$

$$+ \frac{9}{4}\log(2\pi e) + \frac{9}{2}\log\mathsf{E}\left[\|\hat{\mathbf{A}}_1 + \tilde{\mathbf{A}}_1\|^2\right] + \epsilon(n), \tag{48}$$

$$R_Z < \frac{1}{n}h(Z_1^n|M_Y, \tilde{\mathbf{H}}_1^n, \hat{\mathbf{A}}_1^n, \hat{\mathbf{H}}_1^n) - \frac{1}{n}h(Z_1^n|M_Y, M_Z, \tilde{\mathbf{H}}_1^n, \hat{\mathbf{A}}_1^n, \hat{\mathbf{H}}_1^n) + \epsilon(n). \tag{49}$$

Using the chain rule for differential entropy and the fact that differential entropy cannot decrease if conditioning is removed we have

$$h(Y_1^n|\tilde{\mathbf{A}}_1^n, \hat{\mathbf{A}}_1^n, \hat{\mathbf{H}}_1^n) = \sum_{i=1}^n h(Y_i|Y_1^{i-1}, \tilde{\mathbf{A}}_1^n, \hat{\mathbf{A}}_1^n, \hat{\mathbf{H}}_1^n) \leq \sum_{i=1}^n h(Y_i|\tilde{\mathbf{A}}_i, \hat{\mathbf{A}}_i). \tag{50}$$

The terms on the right hand side of (50) can then be upper bounded using the fact that a Gaussian distribution maximizes differential entropy under a second moment constraint, followed in a second step by using the Cauchy-Schwarz Inequality and by splitting up the expectation over $\hat{\mathbf{A}}_i + \tilde{\mathbf{A}}_i$ into the expectations over the magnitude and phase of $\hat{\mathbf{A}}_i + \tilde{\mathbf{A}}_i$ and applying Jensen's Inequality to the expectation over the phase. Further, Lemma 6 can be applied since $\mathsf{E}[\|\mathbf{A}_i\|^2] < \infty$ for every $i$ due to Condition (8) and by defining $\mathcal{E}_i \triangleq \mathsf{E}[\|\mathbf{X}_i\|^2]$ which is finite according to (6). The last inequality follows by using again Jensen's inequality and because the distribution of $\hat{\mathbf{A}}_i + \tilde{\mathbf{A}}_i$ is assumed not to depend on the index $i$:

$$\sum_{i=1}^n h(Y_i|\tilde{\mathbf{A}}_i, \hat{\mathbf{A}}_i) \leq \sum_{i=1}^n \mathsf{E}\left[\frac{1}{2}\log\left(2\pi e\left(\mathsf{E}\left[\left\|\left(\hat{\mathbf{A}}_i + \tilde{\mathbf{A}}_i\right)^\mathsf{T}\mathbf{X}_i\right\|^2 \bigg|(\hat{\mathbf{A}}_i + \tilde{\mathbf{A}}_i)\right] + \sigma^2\right)\right)\right]$$

$$\leq \sum_{i=1}^n \mathsf{E}\left[\frac{1}{2}\log\left(2\pi e\left(\|\hat{\mathbf{A}}_i + \tilde{\mathbf{A}}_i\|^2 \cdot \mathsf{E}\left[\|\mathbf{X}_i\|^2\big|\|\hat{\mathbf{A}}_i + \tilde{\mathbf{A}}_i\|\right] + \sigma^2\right)\right)\right]$$

$$\leq \sum_{i=1}^n \left\{\frac{1}{2}\log\left(2\pi e\left(\mathsf{E}\left[\|\hat{\mathbf{A}}_i + \tilde{\mathbf{A}}_i\|^2\right]\mathcal{E}_i + \sigma^2\right)\right)\right\} + n\frac{\log e}{e}$$

$$\leq \frac{n}{2}\log\left(2\pi e\left(\mathsf{E}\left[\|\hat{\mathbf{A}}_1 + \tilde{\mathbf{A}}_1\|^2\right]\mathcal{E} + \sigma^2\right)\right) + n\frac{\log e}{e}. \tag{51}$$

Furthermore since the random vector $\mathbf{X}_k$ is determined by the messages $M_Y$, $M_Z$ and by the fading coefficients $\hat{\mathbf{A}}_1^n$ and $\hat{\mathbf{H}}_1^n$ and since the sequence $N_1^{(z)}, \ldots, N_n^{(z)}$ is IID $\mathcal{N}(0, \sigma^2)$ it follows that

$$h(Z_1^n|M_Y, M_Z, \tilde{\mathbf{H}}_1^n, \hat{\mathbf{A}}_1^n, \hat{\mathbf{H}}_1^n) = \frac{n}{2}\log(2\pi e\sigma^2). \tag{52}$$

Combining the bounds (48) and (49) and applying (50), (51) and (52) one obtains

$$R_Y + \frac{1}{2}R_Z < \frac{1}{2}\log\left(1 + \frac{\mathsf{E}\left[\|\hat{\mathbf{A}}_1 + \tilde{\mathbf{A}}_1\|^2\right]\mathcal{E}}{\sigma^2}\right) - \frac{9}{2n}h(\tilde{\mathbf{A}}_1^n|\hat{\mathbf{A}}_1^n, \hat{\mathbf{H}}_1^n) + \gamma'$$

$$+ \frac{1}{2}\mathsf{E}\left[\log^+ \|\hat{\mathbf{H}}_1 + \tilde{\mathbf{H}}_1\|\right] + \mathsf{E}\left[\log^+ \frac{1}{\|\hat{\mathbf{A}}_1 + \tilde{\mathbf{A}}_1\|}\right]$$

$$+ \frac{9}{2}\log\mathsf{E}\left[\|\hat{\mathbf{A}}_1 + \tilde{\mathbf{A}}_1\|^2\right] + \frac{3}{2}\epsilon(n) \tag{53}$$

where $\gamma' = \gamma + \frac{\log e}{e} + \frac{10}{4}\log(2\pi e)$.

By exchanging the roles of the two receiving terminals in all the steps an upper bound on $\frac{1}{2}R_Y + R_Z$ is obtained. This upper bound summed up with Bound (53) and divided by $\frac{3}{2}$ yields

$$\begin{aligned}
R_Y + R_Z < {} & \frac{1}{3}\log\left(1 + \frac{\mathsf{E}\left[\|\hat{\mathbf{A}}_1 + \tilde{\mathbf{A}}_1\|^2\right]\mathcal{E}}{\sigma^2}\right) + \frac{1}{3}\log\left(1 + \frac{\mathsf{E}\left[\|\hat{\mathbf{H}}_1 + \tilde{\mathbf{H}}_1\|^2\right]\mathcal{E}}{\sigma^2}\right) \\
& + \frac{4}{3}\gamma' - \frac{3}{n}h(\tilde{\mathbf{A}}_1^n|\hat{\mathbf{A}}_1^n, \hat{\mathbf{H}}_1^n) - \frac{3}{n}h(\tilde{\mathbf{H}}_1^n|\hat{\mathbf{A}}_1^n, \hat{\mathbf{H}}_1^n) \\
& + \frac{1}{3}\mathsf{E}\left[\log^+\|\hat{\mathbf{A}}_1 + \tilde{\mathbf{A}}_1\|\right] + \frac{1}{3}\mathsf{E}\left[\log^+\|\hat{\mathbf{H}}_1 + \tilde{\mathbf{H}}_1\|\right] \\
& + \frac{2}{3}\mathsf{E}\left[\log^+\frac{1}{\|\hat{\mathbf{A}}_1 + \tilde{\mathbf{A}}_1\|}\right] + \frac{2}{3}\mathsf{E}\left[\log^+\frac{1}{\|\hat{\mathbf{H}}_1 + \tilde{\mathbf{H}}_1\|}\right] \\
& + 3\log\mathsf{E}\left[\|\hat{\mathbf{A}}_1 + \tilde{\mathbf{A}}_1\|^2\right] + 3\log\mathsf{E}\left[\|\hat{\mathbf{H}}_1 + \tilde{\mathbf{H}}_1\|^2\right] + 2\epsilon(n).
\end{aligned} \quad (54)$$

Taking now the limit as $n$ tends to infinity the term $\epsilon(n)$ tends to 0 and the sum rate can be upper bounded by

$$\begin{aligned}
R_Y + R_Z < {} & \frac{1}{3}\log\left(1 + \frac{\mathsf{E}\left[\|\hat{\mathbf{A}}_1 + \tilde{\mathbf{A}}_1\|^2\right]\mathcal{E}}{\sigma^2}\right) + \frac{1}{3}\log\left(1 + \frac{\mathsf{E}\left[\|\hat{\mathbf{H}}_1 + \tilde{\mathbf{H}}_1\|^2\right]\mathcal{E}}{\sigma^2}\right) \\
& + \frac{4}{3}\gamma' - 3\lim_{n\to\infty}\frac{1}{n}h\left(\tilde{\mathbf{A}}_1^n \,\middle|\, \hat{\mathbf{A}}_1^n, \hat{\mathbf{H}}_1^n\right) - 3\lim_{n\to\infty}\frac{1}{n}h\left(\tilde{\mathbf{H}}_1^n \,\middle|\, \hat{\mathbf{A}}_1^n, \hat{\mathbf{H}}_1^n\right) \\
& + \frac{1}{3}\mathsf{E}\left[\log^+\|\hat{\mathbf{A}}_1 + \tilde{\mathbf{A}}_1\|\right] + \frac{1}{3}\mathsf{E}\left[\log^+\|\hat{\mathbf{H}}_1 + \tilde{\mathbf{H}}_1\|\right] \\
& + \frac{2}{3}\mathsf{E}\left[\log^+\frac{1}{\|\hat{\mathbf{A}}_1 + \tilde{\mathbf{A}}_1\|}\right] + \frac{2}{3}\mathsf{E}\left[\log^+\frac{1}{\|\hat{\mathbf{H}}_1 + \tilde{\mathbf{H}}_1\|}\right] \\
& + 3\log\mathsf{E}\left[\|\hat{\mathbf{A}}_1 + \tilde{\mathbf{A}}_1\|^2\right] + 3\log\mathsf{E}\left[\|\hat{\mathbf{H}}_1 + \tilde{\mathbf{H}}_1\|^2\right]
\end{aligned} \quad (55)$$

Theorem 1 follows then by taking the limit of the ratio of the right hand side of (55) to $\log\left(1 + \frac{\mathcal{E}}{\sigma^2}\right)$ as $\mathcal{E}$ tends to infinity since due to Assumptions (8), (9) and (10) all the additive terms except for the first two in expression (55) are bounded for every value of $\mathcal{E}$. In particular the terms on the second last line are finite since conditioning on $\hat{\mathbf{A}}_1$ and $\hat{\mathbf{H}}_1$, respectively, [10, Lemma 6.7 c)] shows that the terms are finite if $h(\tilde{\mathbf{A}}_1|\hat{\mathbf{A}}_1)$ and $h(\tilde{\mathbf{H}}_1|\hat{\mathbf{H}}_1)$ are finite which follows by Assumptions (9) and (10). Furthermore it can be showed using Jensen' Inequality that the terms $\mathsf{E}\left[\log^+\|\hat{\mathbf{H}}_1 + \tilde{\mathbf{H}}_1\|\right]$ and $\mathsf{E}\left[\log^+\|\hat{\mathbf{A}}_1 + \tilde{\mathbf{A}}_1\|\right]$ stay finite whenever $\mathsf{E}[\|\mathbf{A}_1\|^2]$ and $\mathsf{E}[\|\mathbf{H}_1\|^2]$ are finite which holds due to (8).